\documentclass[a4paper]{jpconf}
\usepackage{graphicx}
\newcommand{\be}{\begin{eqnarray}}
\newcommand{\ee}{\end{eqnarray}}
\begin{document}
\title{Conical Flow induced by Quenched QCD Jets}

\author{ J.Casalderrey-Solana,~E.V. Shuryak and D.Teaney}

\address{ Department of Physics and Astronomy\\ State University of New York,
     Stony Brook, NY 11794-3800}

\ead{casalder@tonic.physics.sunysb.edu, shuryak@tonic.physics.sunysb.edu, 
     dteaney@tonic.physics.sunysb.edu}

\begin{abstract}
Quenching is a recently discovered phenomenon in which QCD jets
created in heavy ion collisions deposit a large fraction or even all
their  energy and momentum into the produced matter . 
At RHIC and higher energies, where that matter is a 
 strongly coupled Quark-Gluon Plasma (sQGP) with
very small viscosity, we suggest that this energy/momentum  propagate 
as a collective excitation or ``conical flow''.
Similar hydrodynamical phenomena are well known, e.g. the
so called sonic booms from supersonic planes.
We solve the linearized relativistic 
hydrodynamic equations to detail the flow picture. We argue that
for RHIC collisions the direction of this flow should make a cone at
a specific large angle with the jet, of about $70^o$, and thus lead
to peaks in particle correlations at the angle $\Delta\phi=\pi\pm 1.2$
rad relative to the large-$p_t$ trigger. This
angle happens to match 
perfectly 
the position of the maximum in
the angular
distribution of secondaries associated with the trigger recently
seen by the STAR and PHENIX collaborations. We also discuss briefly
possible alternative explanations and suggest some further tests
to clarify the mechanism.
\end{abstract}

\section{Introduction}
{\bf Jet quenching} is an important phenomenon, predicted in a number
of papers \cite{early} and recently observed at RHIC
\cite{jetquenching}. The main  research has been so far related to 
the calculation of the energy loss $dE/dx$ of the fastest
parton, and focused on
a 
kind of tomography of the produced matter.
 In this work, however, we focus on a different question: {\em Where
   does the energy of the quenched jets go?  } 
We
point out that
as the hydrodynamical
description of sQGP excitations works well, it should be used to
predict how
collective flow would develop, after a local deposition
of energy and momentum. As we will see below,
it will be a coherent source of sound waves in conical form.

In what follows, we would like
 to treat two different types of energy losses separately :
(i) the {\em radiative} losses, producing relativistic gluons
in the forward direction;
 and
(ii) the {\em scattering/ionization} losses, which deposit energy
and momentum directly
 into
the medium, as well as radiative losses of gluons at rather large
angles
(see Discussion section).
Such gluons are rather soft and are promptly absorbed by the medium.

The study of radiative losses
were started in \cite{Gyulassy_losses},  then corrected 
for the destructive interferences (the so-called Landau-Pomeranchuck-Migdal
or LPM  effect) in~\cite{Dok_etal}. For a recent brief 
summary, see also~\cite{xnwang_workshop}. Although it is  
the dominant mechanism for
quenching hard ($z\sim1$) fragmenting hadrons, in our problem 
 the second type of losses are more important because
  the primary parton and
the radiated
gluons all move with speed close to the speed of  light and are 
treated as one object.

Elastic energy losses were first studied by Bjorken\cite{early}, while
those due to 
``ionization'' of bound states in sQGP were recently considered by Shuryak
and Zahed \cite{SZ_dedx}. These mechanism deposit additional
  energy, momentum and
entropy into
the matter.  (Like for delta electrons in ordinary matter, 
this excitation kicks particles mostly orthogonal
to the jet direction.)  
It is their combined magnitude,  $dE/dx= 2-3 \, GeV/fm$, the one we will
use below.  Even at such loss rates, a
jet
passing through the diameter of the fireball, created in central Au-Au
collisions, may deposit
up to 20-30 GeV, enough to absorb the jets
of interest at RHIC.
 
Let us start our discussion 
of  associated collective effects 
by recalling
  {\bf the energy scales involved.}
While the total CM energy in a Au-Au collision at RHIC
is very large (about 40 TeV) compared to the energy of a
 jet (typically 5-20 $\, GeV$), the jet energy is transverse.
The total transverse 
energy of all secondaries per one unit of rapidity is $dE_\perp/dy\sim 600\,
GeV$. Most of it is thermal, with only about
100 GeV being
related to collective motion.  Furthermore, the so called  elliptic flow
 is a $\sim 1/10$ asymmetry  and therefore it carries energy 
$\sim 10 \,GeV$ which is 
comparable to that lost by jets. 
 Since elliptic flow was observed and studied in detail, we conclude that 
 conical
flow should be observable as well. (In order to separate the two,
it is  beneficial to focus first on the most central collisions,
 where the
 elliptic flow is as small as possible.)
\begin{figure}
\begin{center}
 \includegraphics[width=7cm]{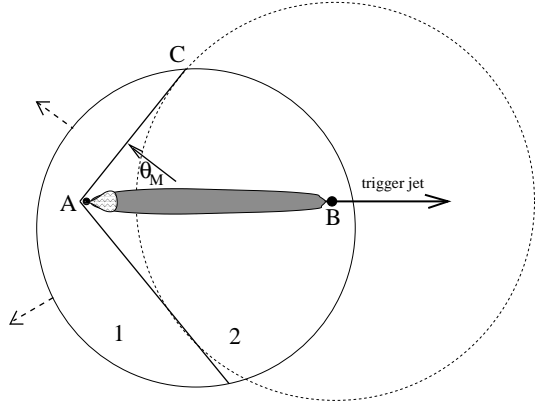}
 \caption{\label{fig_shocks}
A schematic picture of flow created by a jet going through
the fireball. The trigger jet is going to the right
from the origination point (the black circle at point B)
from which sound waves start propagating as spherical waves (the
dashed circle). The
  companion quenched jet is moving to the left, heating the matter
and thus creating a cylinder of additional matter (shaded area).
 The head of the jet is a ``nonhydrodynamical core'' of the QCD gluonic shower,
formed by the original hard parton (black dot A).
The solid arrow shows a direction of flow normal to the shock cone at
the angle $\theta_M$, the 
dashed arrows show the direction of the flow after the shocks hit the edge
of the fireball.
}
\end{center}
 \end{figure}

  Fig.\ref{fig_shocks} explains a view of the process
in a plane transverse to the beam.  Two oppositely
moving jets originate from   the hard collision point B.
 Due to strong quenching, the survival of the trigger
jet biases it to be produced close to the surface and to
 move outward. This  forces its companion to 
move inward through matter and to be maximally quenched.
The energy deposition starts at point B, thus a spherical sound wave
appears (the dashed circle in Fig.\ref{fig_shocks} ). Further 
 energy deposition is along the jet line, and is propagating at the speed of
light, 
till the leading parton is found at point A
at the moment of the snapshot.

As  is well known, the interference of perturbations from  
 a supersonically moving body 
(such as a supersonic jet plane or a meteorite)
creates a conical flow behind the
 shock waves. Similar flow was discussed in Refs\cite{old_conical}
for shocks in cold nuclear matter, in which compression up to 
QGP production takes place. Unfortunately,
experiments have shown that nuclear matter is
 too
dilute and dissipative to make shocks: but 
discoveries made at RHIC 
 let us be more optimistic in the sQGP case we consider now \cite{fnStocker}.

The angle $\theta_M$ defined in the figure 
 is given by a simple geometrical
condition:  the distance
traveled by the jet during the proper time interval $\tau$
is  $AB=c\tau$, while the one
traveled by the wave is
\be 
\label{eqn_Mach}
CB=\int_0^\tau c_s(t)dt ~~ \Longrightarrow ~~ 
cos\theta_M={1\over c\tau}\int_0^\tau c_s dt ~.
\ee 
We use the speed of sound since,
 apart from the region close to the
head of the object
\cite{fncs}, the shock waves are weak and thus they move with the
speed of sound
(we use units in which c=1.)

The region near the head of the jet,
which we will refer to as a ``non-hydrodynamical core'',
will
not be discussed  in this work. Let us just mention that
near point B, where the jet is produced,
it consists of an ``undressed'' hard parton only. 
However, as it is constantly
emitting gluons, which emit new ones  etc., the whole shower is a
complicated nonlinear phenomenon which should obviously be treated
via the tools of quantum field theory \cite{fnRM}.
As found in \cite{Dok_etal},
the multiplicity of
this shower grows nonlinearly with time, so eventually
 the core may become a  macroscopic body, providing a large
perturbation of the matter. From the hydrodynamical point of view, its
size is limited from below by the dissipative ``sound attenuation length''
$\Gamma_s=(4/3)\eta/(\epsilon+p)$, with $\eta$ being the shear viscosity.

A  shaded region in 
Fig.\ref{fig_shocks} consists of ``new matter'' related to the entropy 
produced in the process $ dS/dx$, which can be calculated only with dissipative
dynamics in the near zone, which we do not attempt in this work.
 Matter is expected to get equilibrated soon, 
 and thus the radius of the cylinder $R_c$ of new matter can be estimated as
$ dS/dx = s(T) \pi R_c^2 $.
As is well known, 
a constant size cylinder does not emit any sound.

\section{Linearized Hydrodynamics}
The hydrodynamical equations we use are  simpler than those used
in   Refs\cite{old_conical} since we will consider that the total 
energy-momentum density deposited by the 
jets
is a small perturbation compared to the total energy of the medium. This allows us to linearize the problem. The approximation breaks down
close to the jet, where we will not describe the hydrodynamic fields.
 We will use cylindrical coordinates with $x$ and $r$ parallel and
perpendicular to the jet axis, respectively.
We will also assume that the perturbed medium
is homogeneous and at rest.

In the linearized approximation we define the following quantities 
in terms of perturbations of the stress energy tensor:
\be
\epsilon=\delta  T^{00} ~~,~~
g^i=\delta   T^{0i} ~.
\ee
 The remaining non-zero components of the stress tensor \cite{Derek_visc} are:
\be
T^{ij}=c^2_s \epsilon \delta^{ij}-\frac{\eta}{\epsilon_0+p_0} \left< \partial^i g^j \right > ~~,~~
 \left \langle \partial^i g^j \right \rangle=\partial^i g^j+\partial^j g^i-\frac{2}{3}\delta^{ij}\partial_i g^i ~.
\ee
Here we have used $\frac{\partial p}{\partial e} = c^2_s  $ and recall that in 
the linearized approximation the velocity field is 
$v^i=\frac{g^i}{\epsilon_0+p_0}$; $\epsilon_0$ and $p_0$ are the energy 
density and pressure of the unperturbed medium.  

The energy and momentum conservation equations
 $\partial_\mu \delta T^{\mu \nu}$, can be written in Fourier space, where we observe
that if we define $\vec{g}=g_{\scriptstyle \rm L}\frac{\vec{k}}{k}+\vec{g_{\scriptstyle \rm T}}$ (where L ant T stand
for longitudinal and transverse respectively), the linearized hydro equation 
decouples as:
\be
\label{eglsys}
\partial_t \epsilon+i k g_{\scriptstyle \rm L}=0 ~,\nonumber   \\ 
\partial_t g_l+ic^2_s k_{\scriptstyle \rm L}\epsilon+
\frac{\eta}{\epsilon_0+p_0}\frac{4}{3}k^2g_{\scriptstyle \rm L}=0 ~,
\\
\label{trans}
\partial_t \vec{g_{\scriptstyle \rm T}} + \frac{\eta}{\epsilon_0+p_0}k^2\vec{g_{\scriptstyle \rm T}}=0 ~. 
\ee
The system of equations (\ref{eglsys}) describes sound waves (propagating modes). Equation 
(\ref{trans}) is the diffusion equation and is not propagating.
Only the sound waves will form the Mach cone.

\section{Initial conditions}
 The initial conditions  are set by
the process of thermalization of the energy and momentum lost by the jet. This 
complex
process should take place at distances of 
order $\Gamma_s$ from the production point. As $\Gamma_s$ is also the 
minimal size of the liquid cells, we will simply consider that there is a
variation of $T^{\mu\nu}$ at the position of the particle. 

If we consider an infinitesimal displacement $dt_0$ of the high energy particle (moving 
with the speed of light) in the $x$ direction at the point $(t_0,t_0,\vec{0})$, we need to specify 
the infinitesimal variation of the fields $\epsilon_{dt_0}(t=t_0,\vec{x})$, 
$\vec{g_{dt_0}}(t=t_0,\vec{x})$. Due to the symmetries of the problem, 
the most general expression for those functions is:
\be
\label{ini_cond}
\epsilon_{dt_0}(t=t0,\vec{x})=e_0(z,r) ~~,~~
\vec{g}_{dt_0}(t=t0,\vec{x})=g_0(z,r)\delta^{ix}
+\vec{\partial} g_{1}(z,r) ~.
\ee 
As argued before $e_0$, $g_0$ and $g_{1}$ are some functions with
characteristic scale $\sigma \sim \Gamma_s$. The exact expression for these functions 
depends on the details of the thermalization process.
As we do not know these details, we will need to make 
certain assumptions about the different functions that 
appear in (\ref{ini_cond}).
In general, one may
consider two different scenarios of matter excitation:

\emph{Scenario 1} in which local deposition of energy and momentum
is described by the
first terms in (6),  $e_0$ and $g_0$. This imposes a constraint 
 in 
(\ref{ini_cond}):
\be
\label{total_EP}
\int d^3x \epsilon_{dt_0}(t=t0,\vec{x})=\frac{dE}{dx}dt_0 ~~,~~
\int d^3x g^x{dt_0}(t=0,\vec{x})=\frac{dP}{dx}dt_0 ~,
\ee
where $\frac{dE}{dx}$ and $\frac{dP}{dx}$ are 
 an input. In this case  
 $g_1=0$.
These simple initial conditions, however, excite the ``diffuson mode''
and thus be discarded later.

\emph{Scenario 2} in which the excitation is
due to the gradient term  $g_1$ in (6), with $\epsilon_0=0$ and $\vec{g_0}=0$.
The deposition of energy and momentum in this case comes from
the second order effects. The normalization of such solutions are better done
at large distances, via the energy flow and the
momentum
flow through a large cylinder. As we show below, this generates
the conical solution with sound excitation.

The empirical fact that the second scenario describes the data and
the first one does not obviously provides some insights about
the excitation mechanisms, but we would not speculate about them
now.

\section{Solutions}
In order to solve (\ref{eglsys}) and (\ref{trans}) with the 
initial condition (\ref{ini_cond}), 
we will calculate the corresponding kernels, 
 that is, we will set the functions appearing in (\ref{ini_cond}) to be 
$\delta$-functions. Following the standard procedure, we convolute these 
kernels with some parametrization for $e_0$, $g_0$, and $g_1$:
\be
T^{0\mu}(x)=\int d^3y \left(K^{\mu}_{e}\left(x-y\right) e_0(y)+
                        K^{\mu}_{g_0}\left(x-y\right) g_0(y)+
                        K^{\mu}_{g_1}\left(x-y\right) g_1(y)\right) ~.
\ee
 and integrate over the trajectory of the particle.

\emph{Scenario 1.} In this case we only consider point like excitation in $e_0$, $g_0$.  Keeping terms to order $O(\Gamma_s^2)$ we find the following 4-vector
kernels:

\be
\label{kere}
K^{\mu}_{e}=\sum_{i=\pm}
                   \left( \frac{-1}{4\pi r} \partial_r P_i(\Gamma_s),
                   \partial^k\frac{c_s}{4\pi r}i P_i(\Gamma_s)\right) ~, 
\ee
\be
\label{kerg}
K^{\mu}_{g_0}=\sum_{i=\pm}
                    \left(\partial_x \frac{iP_i(\Gamma_s)}{4\pi c_sr} ,
                    \partial^k\partial_x\frac{-\int^r_0 P_i(\Gamma_s)}
                                                          {4\pi r}\right)
		    +\left(0,\partial^k\partial_x\frac{2 \int^r_0 P(\frac{3}{2}\Gamma_s) }{4\pi r} 
                       +
             \delta^{kx}\frac{ e^{\frac{-r^2}{2\frac{3}{2}\Gamma_s t}}}{\left(2\pi\frac{3}{2}\Gamma_s t\right)^{3/2}}
             \right)
~.
\ee
Where
 $r=|\vec{x}-\vec{x}(t_0)|$ is
the distance from the observation point to the jet position and we have defined
the functions:
\be
P_{\pm}(\Gamma_s)=\frac{1}{\sqrt{2\pi\Gamma_st}}e^{-\frac{(r\pm c_st)^2}{2\Gamma_s t}} ~~,~~
P(\frac{3}{2}\Gamma_s)=\frac{1}{\sqrt{2\pi\frac{3}{2}\Gamma_st}}e^{-\frac{r^2}{2\frac{3}{2}\Gamma_s t}} ~. 
\ee
The second term in (\ref{kerg})
is the (Fourier transform of the) so called diffusion mode (or diffuson) (\ref{trans}). 
It creates a dissipating flow of matter comoving with the jet that 
remembers the location of deposition of momentum. As we will show below, 
this type of excitation leads to a forward peak in the final spectrum   
filling up the Mach cone 
(related to 
(\ref{kere}) and the first term in (\ref{kerg})).

These kernels should be convoluted with the initial distributions. We take a 
simple parametrization (Gaussian) that fulfills the requirement 
(\ref{total_EP}):
\be
e_0(t=t_0,\vec{x})=
    \frac{dE}{dt_0}dt_0\frac{e^{-\frac{(\vec{x}-t_0\hat{x})^2}{2\sigma^2}}}{(2\pi\sigma^2)^{3/2}}
~~,~~ 
g_{0}(t=t_0,\vec{x})=
    \frac{dP^x}{dt_0}dt_0\frac{e^{-\frac{(\vec{x}-t_0\hat{x})^2}{2\sigma^2}}}{(2\pi\sigma^2)^{3/2}} \delta^{kx}
~.
\ee

\emph{Scenario 2} Following the same procedure 
we find the 4-kernel:
\be
K^{\mu}_{g_1}=\sum_{i=\pm}
                   \left( \frac{i}{4\pi c_s r} \partial^2_r P_i(\Gamma_s),
                   \partial^k\frac{-1}{4\pi r}\partial_r P_i(\Gamma_s)\right) 
~,
\ee
where the definitions are the same as in (\ref{kere}) and (\ref{kerg}). 
Note that as claimed, we now find a propagating
solution, the sound wave.

  For example, let us
 assume a simple parametrization for the initial source:
\be
g_{1}(t=t_0,\vec{x})\sim 
e^{-\frac{(\vec{x}-t_0\hat{x})^2}{2\sigma^2}} ~.
\ee

\begin{figure}
\begin{center}
\includegraphics[width=7cm]{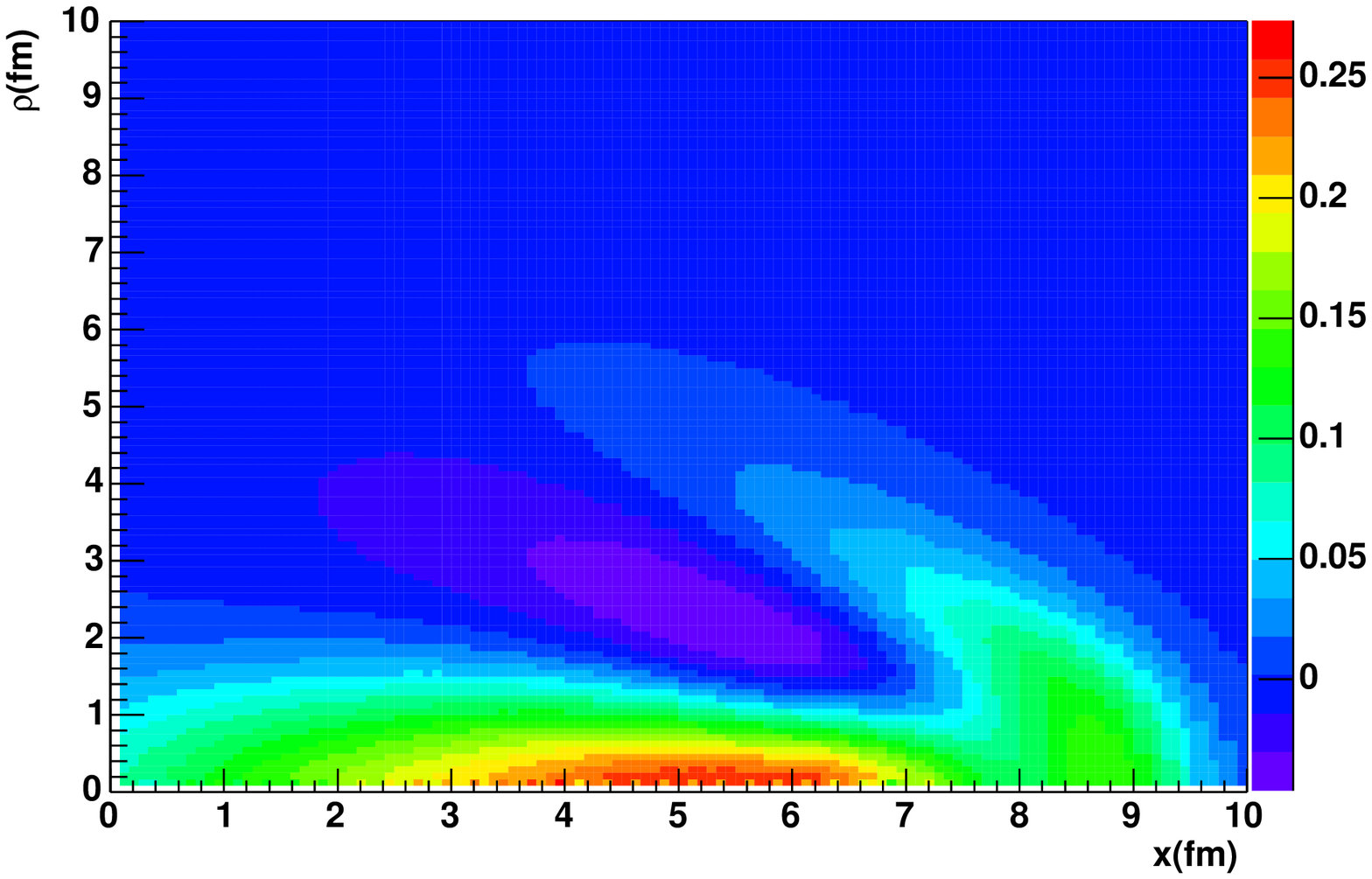}
\includegraphics[width=7cm]{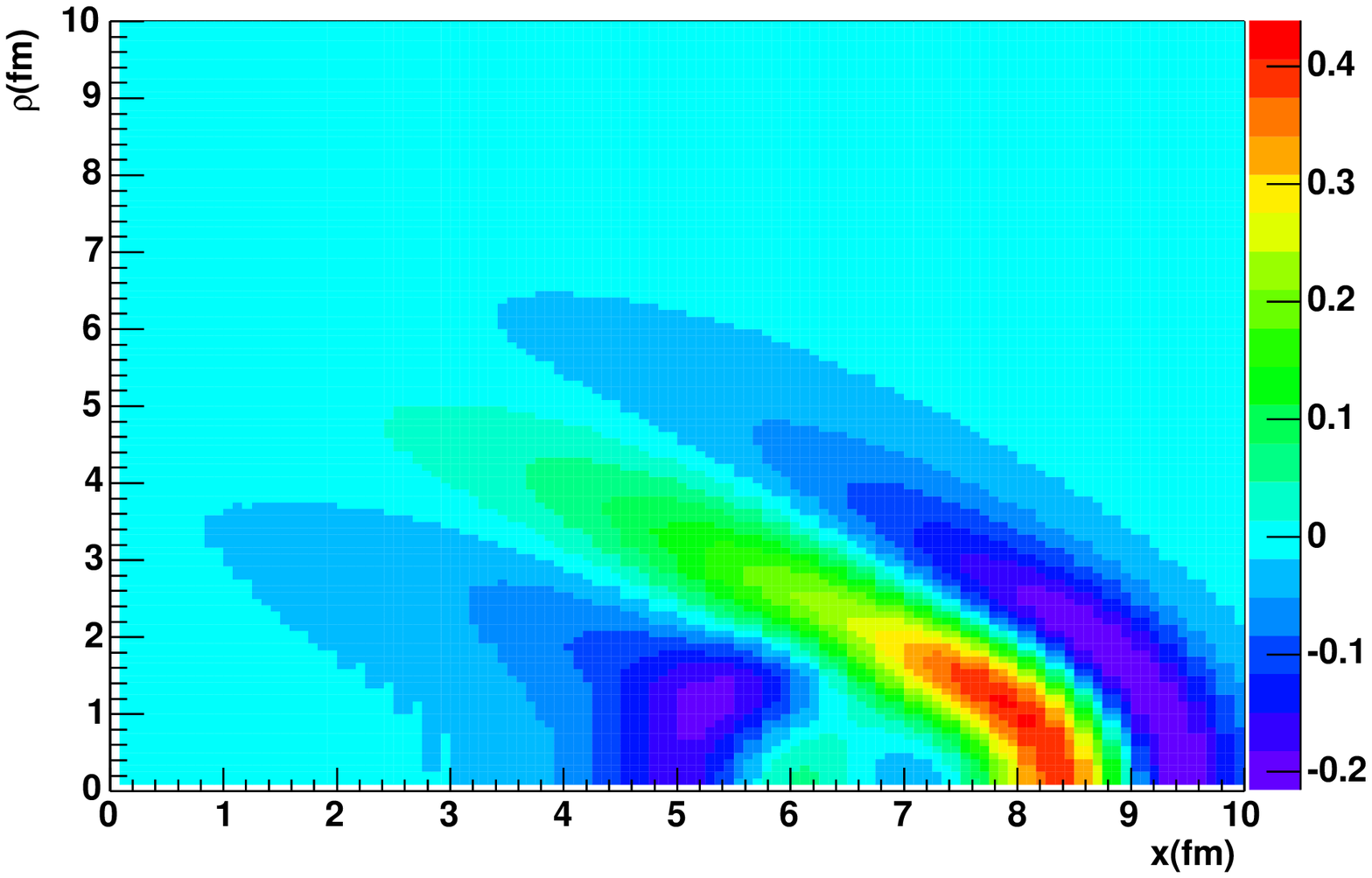}
 \caption{\label{fig_vx}
Velocity field $v_x$ created by a jet moving along the $\hat{x}$ direction
($c^2_s=1/3$, $\Gamma_s=1/(4\pi T)$, $\sigma=\Gamma_s$). The jet is
assumed to disappear at $t=7$ fm while the spectrum calculated at
$t=10$ fm. The two figures (a) and (b) are for scenarios 1 and 2, respectively.
 The values of the parameters are arbitrary. Note that in (a) matter moves preferentially along the 
$\hat{x}$ direction, while in (b) it is in the Mach direction (\ref{eqn_Mach}).
 }
\end{center}
 \end{figure}

In Fig.\ref{fig_vx} we show the velocity field along the jet axis generated 
by the high energy particles according to both models of the initial 
conditions. In Fig. \ref{fig_vx}(a) we can observe the effect of the diffuson
as a maximum in the velocity along the $\hat{x}$ axis, where the jet 
propagates. In Fig. \ref{fig_vx}(b) such maximum does not exist, as there is no
excitation of the diffusion mode. We also observe in both cases the sound wave,
that is more prominent in the second case. Note also
that in both cases we have positive and negative velocities, that is, matter
moving outward and inward (as it should be due to conservation of matter).

\section{Spectrum} 
\begin{figure}
\begin{center} 
\includegraphics[width=7cm]{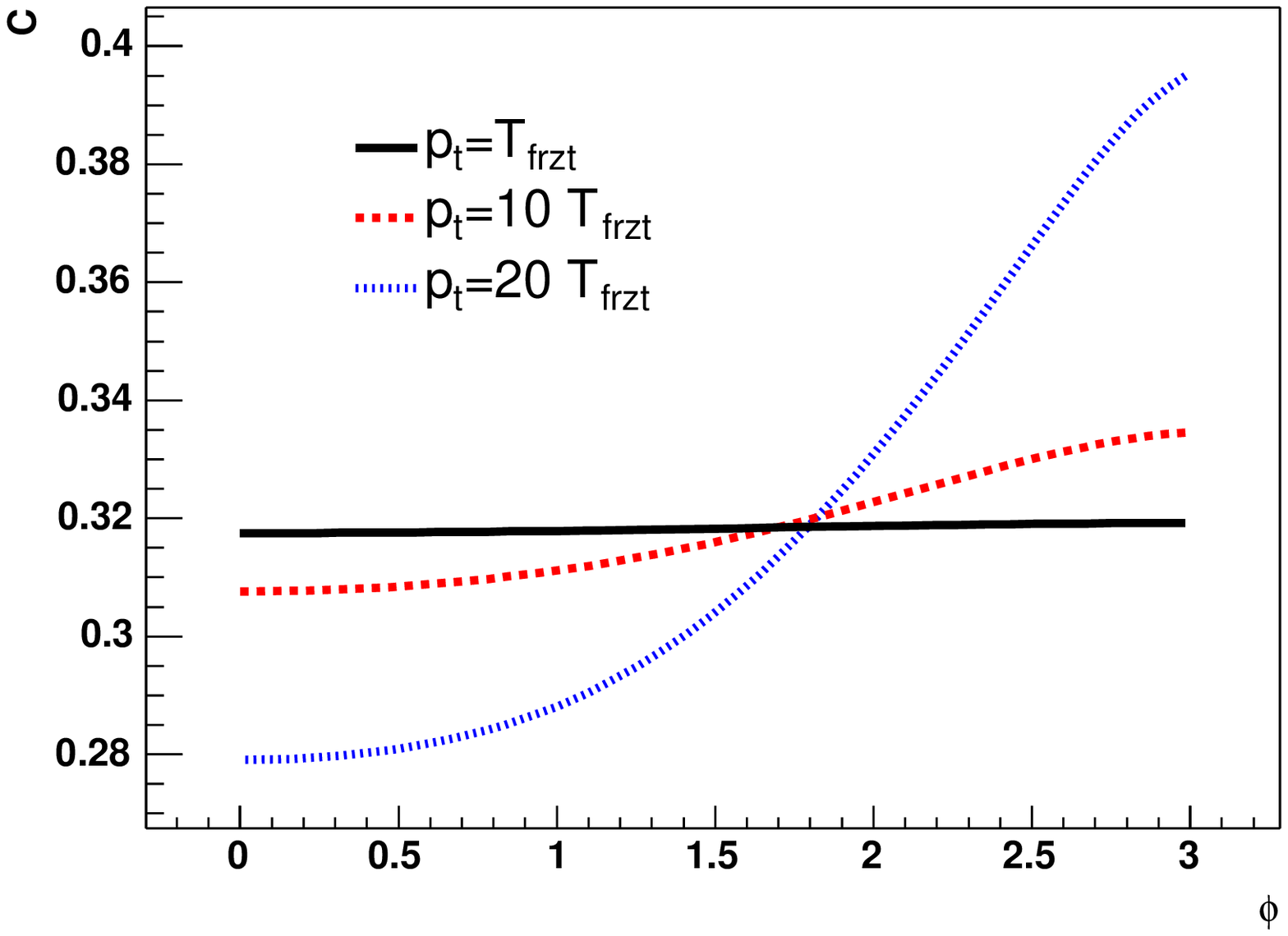}
\includegraphics[width=7cm]{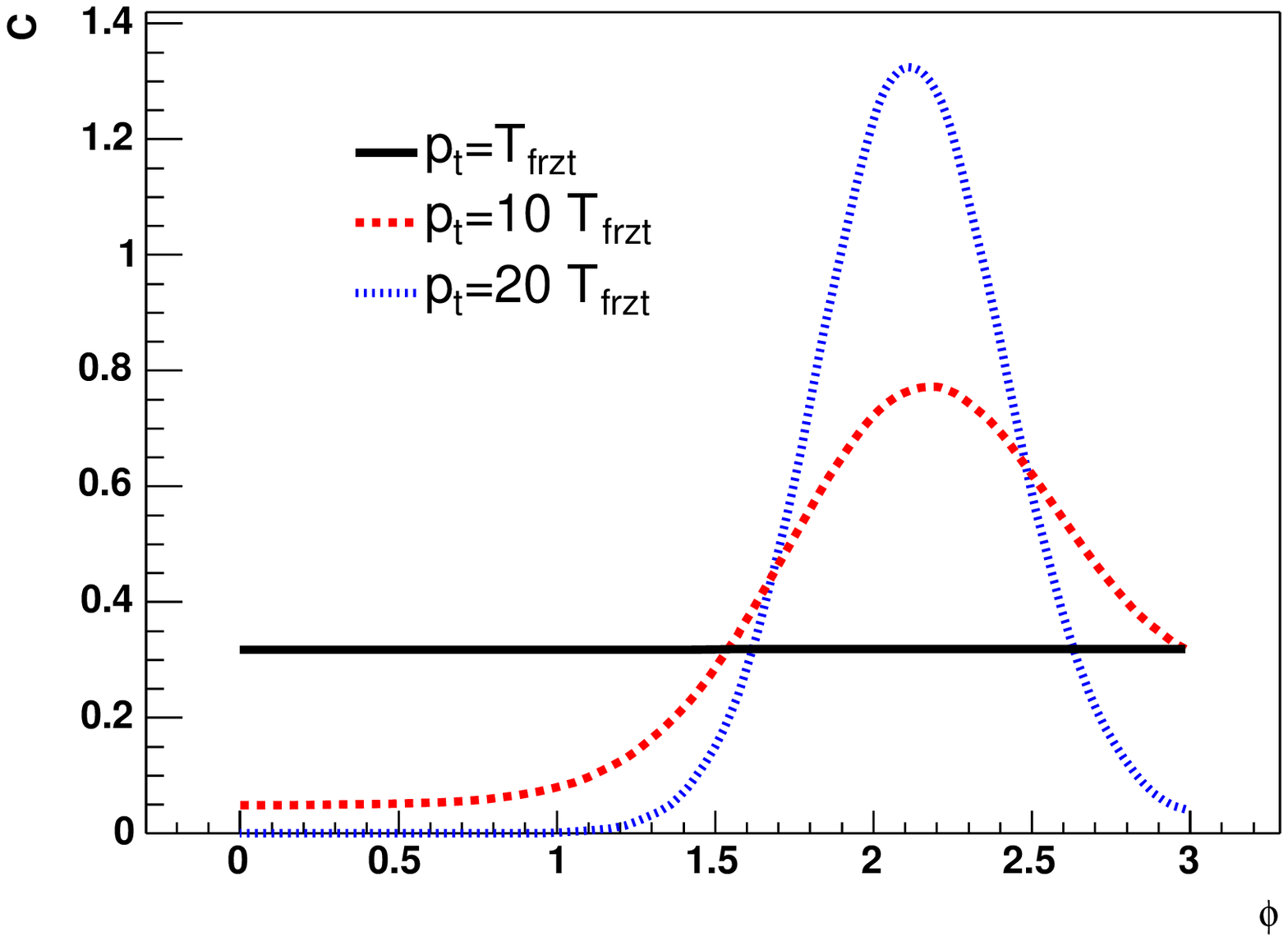}
\end{center}
 \caption{\label{spectrum}
The normalized spectrum of associated secondaries versus the azimuthal
angle $\phi$. Three curves are for
different $p_t$ at $y=0$ for 
$c^2_s=1/3$, $\Gamma_s=1/(4\pi T)$, $\sigma=\Gamma_s$. Note the
different scales. The jet disappears completely
 at $t=7$ fm while the spectrum is calculated at  $t=10$ fm. The two figures (a)
and (b) are for scenarios 1 and 2, respectively. 
}
 \end{figure}

We use the previous hydrodynamic fields to calculate the final
spectrum induced by the jet. To do so, we use the standard Cooper-Fry prescription. As our initial
medium is static, we use fixed time freeze-out (neglecting the effect
of the perturbation on the freeze out):
\be
\label{mspectrum}
\frac{dN}{d^3p}=\int_V d^3V e^{-\frac{E}{T_{frzt}}+\delta} ~~,~~
\delta= \frac{E}{T_{frzt}} \frac{\delta T}{T_{frzt}} + \frac{\vec{p}\vec{v}}{T_{frzt}} 
~,
\ee
where V is the volume of the fireball and $T_{frzt}$ is the freeze-out  
temperature. According to our approximation, we expand the exponent
to first order in the perturbation. We can now
discuss two opposite regimes
in the spectrum:

{\bf Low energy particles $E\sim T_{frzt}$}. In this region 
we can expand the exponential in (\ref{spectrum})
and express the spectrum in terms of the energy and momentum deposited 
(to first order):
\be
\label{lowspc}
\frac{dN}{d^3p}=e^{-\frac{E}{T_{frzt}}}\left(V+\frac{E}{T_{frzt}}
                \frac{E_{dep}}{\epsilon_0+p_0} 
                +\frac{\vec{P}}{T_{frzt}}
                  \frac{\vec{P}_{dep}}{\epsilon_0+p_0}\right)
~.
\ee          
One finds that soft particles are insensitive to the particular shape
                of the flow field and their
 angular dependence  is just a cosine
of the relative angles of the observed particle and the jet.

{\bf High energy particles $E >> T_{frzt}$} which
compensates the smallness of the flow velocity. The integral is now
 dominated by the 
maximum of the exponent. Thus, only the points of maximum modifications of 
hydrodynamic fields contribute to the final spectrum. It is clear then that
in the scenario 1 the  diffusion
mode totally dominates the spectrum and there are no angular correlations
related to the speed of sound. However, in the second scenario
 one does
not excite that mode, and the final spectrum do reflect the shape of the
sonic disturbance.

In order to illustrate the effect of the modification of the final 
particle production due to a jet moving with $y=0$ and $\phi=\pi$, we show in Fig.\ref{spectrum} the normalized spectrum
defined as follows:
\be
C=\frac{1}{Q} \frac{dN}{dyd^2p_t}(y=0,p_t,\phi) ~~,~~ 
Q=\int^{2\pi}_{0} d\phi\frac{dN}{dyd^2p_t}(y=0,p_t,\phi)  ~.
\ee
We observe that, as claimed, the effect of the diffusion mode hides the
Mach cone in the final production. As a consequence, we do not observe such a
structure in fig Fig.\ref{spectrum}(a); we do observe it, however, in Fig.\ref{spectrum}(b), where such mode is not excited. 

The parameters in Fig.\ref{spectrum} are arbitrary (with the  exception of $\Gamma_s$,
which is set to its minimal bound \cite{visc_bound}), and are chosen as a matter of illustration.
Let us remark that these spectra are very sensitive to the value of $\Gamma_s$,
which could provide a experimental constraint on its value, provided a correct
understanding of the deposition process.

\section{How can the effect be observed?}
The most important feature of conical flow is  
 its direction, which is normal to the 
shock front and thus making the so called Mach 
angle  (\ref{eqn_Mach}) with the jet direction,
determined basically by the speed of sound in matter.
So one can conclude that quenched jets must be accompanied
by a cone of particles with the opening angle $\theta_M$. 
This cone angle should be the same for any jet energy (in
contrast to radiation angles, which are shrinking at high energy).

  The
appropriate value for the speed of sound is a time-weighted
average of three stages: (i) the QGP phase ($c_s\approx 1/\sqrt{3}$), (ii) the mixed phase
$c_s\approx 0$ and (iii)
the hadronic or ``resonance gas'' stage, with
 $c_s\approx \sqrt{.2}$ \cite{res_gas}. For RHIC we found the 
time-weighted average
to be $\bar c_s\approx .33$.
 We thus conclude that in this case the emission angle
of the conical flow
should be at angles about 70 degrees relative to the jet, or in radians
\be \theta_{emission} =\pi\pm arccos(\bar c_s/c)\approx 1.9,4.3.\ee

 The first measurements
of interest to this issue
have been made by the STAR collaboration \cite{STAR_peaks}, \cite{MIT_workshop_STAR} 
 which studied two particle correlations
 in which
the trigger particle has $4<p_t<6 ~GeV$, while the associated 
particles 
have basically all momenta (in fact 
$0.15<p_t<4 ~GeV$
).
At $\Delta\phi=0$ one finds a peak
due to particles from the triggered 
jet,
while
particles from the companion jet should result in a peak 
at 
$\Delta\phi=\pi$.
Such peak is clearly seen in pp collisions, in which no matter is present:
but in Au-Au collision one finds instead a {\em double-peaked} distribution
with a $minimum$ at the original jet direction $\Delta\phi=\pi$.
Remarkably
this value $\pi\pm 1.2$ nicely
agrees with  the observed positions of the two
maxima of the distribution of secondaries.

After submission of our paper in the preprint form, two more
important observations have been made, both discussed
at this workshop \cite{MIT_workshop_STAR},\cite{MIT_workshop_PHENIX}.

The PHENIX 
collaboration 
have shown \cite{MIT_workshop_PHENIX} correlation function
 in which the companion particle is harder
than the average $p_t>1\, GeV$.
They also reported data for 
6 centrality classes and in 
all 
 of them
(but the most peripheral one) there is a minimum at 180$^o$
and even sharper peaks at the
same angle as we predict. 
We also remark that our results in Fig.3(b) for $p_t=10 T_{frzt}$ 
($T_{frzt}=100$ MeV) are similar to the correlation functions in \cite{MIT_workshop_PHENIX}.

The STAR collaboration also 
reported \cite{MIT_workshop_STAR}
the associated particle
mean $\left<p_t\right>(\phi)$, which also displays maxima
away from $\phi=\pi$, where  one finds a
clear minimum. This fact strongly supports 
our explanation of the peaks by conical flow.  The fact that at 180$^o$ the mean $p_t$
of associated secondaries is consistent in magnitude with the mean transverse
 momentum in the average background events means that
 in the experimental
 conditions the inward-moving jet  is
completely quenched by freeze-out and  no 
contribution from the ``nonhydro core'' is visible in the data.

 Unfortunately, we can only see a projection on $\Delta\phi$
because the  distribution in rapidity of the associated jet is very
wide,
wiping out another projection of the cone. One could think that 
this wide rapidity distribution could
 also wipe out the double peak structure in the final two particle 
correlation. Even though the full hydrodynamic problem should be solved, 
we can give here an 
argument by which this does not happen. For that we assume boost invariance
of the medium and that the rapidity ($y_j$) distribution of the away jet is flat. 
We will also assume that particles are formed in a cone around the jet axis. It
is clear then, that in a frame with rapidity $y_j$ (where the away side jet is transverse) the spectrum of those particles can only depend
on the momentum of the particle ($p_j$) and its angle with respect to the jet ($\theta^*$). We assume
also that this dependence factorizes and that the dependence in $\theta^*$ is a very
narrow function of angle $\theta_c^*$ (the cone opening angle). Thus, the 
spectrum of particles produced by a jet with rapidity $y_j$ can be written as,
\be
\label{inv_sp}
E\frac{dN}{d^3p}=P(y_j)dy_jf_p(p_j)\delta(\cos(\theta^*) -\cos(\theta^*_c)) ~,
\ee
where $P(y_j)$ is the probability of finding a jet with such rapidity. We will concentrate now on the spectrum of particles at mid rapidity (y=0)
in the 
lab frame. Boosting back to this frame, we find $p_j=p_t \cosh(y_j)$ and 
$\cos(\phi)=\cos(\theta^*)\cosh(y_j)$, where $p_t$ and $\phi$ are the 
transverse momentum and azimuthal angle respectively. Thus:
\be
\delta(\cos(\theta^*) -\cos(\theta^*_c))=
\frac{\delta(y_j -y^*_j)}{\cos(\theta^*_c)\sqrt{1-\frac{\cos^2(\theta^*_c)}{\cos^2(\phi)}}}
 ~~,~~
\cosh(y^*_j)=\frac{\cos(\phi)}{\cos(\theta^*_c)} ~.
\ee
Upon integration on $y_j$ (assuming P constant) we find
\be
E\frac{dN}{d^3p}=Pf_p(p_t \frac{\cos(\theta^*_c)}{\cos(\phi)})\frac{1}{\cos(\theta^*_c)\sqrt{1-
\frac{\cos^2(\theta^*_c)}{\cos^2(\phi)}}} ~,
\ee
which is clearly a peaked distribution around $\phi=\pm \theta^*_c$. What is more, $f_p$ should be a steeply falling function (exponential in our case) of its argument, which
depletes more the fill up. 
A width on the angular dependence will modify the result but a quantitative 
answer to this effect requires further study.

\section{Discussion:} 
We have considered  an idealized case of homogeneous matter at rest.
  The determination of the exact shape of this cone for real
collisions
is not simple.
It is of course just a technical
matter to include the superposition of  radial, elliptic and
conical flows in a single hydro simulation. The open issue, however, would
be the inclusion of viscosity  at the
hadronic stage, which is not supposed to be small. 
Presumably, as for elliptic flow, the use of a
 hadronic cascade afterburner would provide more realistic results.

When the  conical
solution  reaches the surface of the fireball or 
the freeze-out surface,
 the
  energy stored in 
  outward (positive) flow would continue to move in the same
  direction, while the inward (negative) one would be absent.
  As 
 the shock moves into the lower density region near the
 edge,
energy conservation would force its amplitude to grow, like
do the ocean waves  approaching the beach, or the
  whip motion near its thin end. 
 Furthermore,  if the jet is energetic enough to punch through, 
the regions where cones reach the edge
consist of two separated
regions (circular rings if the jet goes via
the diameter)
 moving toward each other
and eventually colliding, where the cone and the fireball surface are
tangent.

We have seen that the choice of initial condition is very important for the
final result, as the excitation of the diffusion modes may hide the cone 
formation in the final spectrum. We have presented two different scenarios
that are, in a sense, opposite limits, as in one of then we do not excite this
mode. The appearance of such mode in the actual initial conditions, will lead
to some (or even complete) fill up of the cone in the two particle correlations.
This fill up, is also expected for high energy particles, in which the jet
punches through. This is however a completely different mechanism from the 
previous, which will lead to a very different $p_t$-spectrum of the associated
particles (exponential in the first, power law in the second) 

Can there be an alternative explanation of the high-angle maxima
observed? Another two different mechanisms have been proposed:

\emph{Large angle gluon radiation}
It has been argued \cite{Vitev} that QCD
radiation calculated with Landau-Pomeranchuck-Migdal effect included,
may have
a maximum at large angles, similar to the Mach angle we discuss. Those
calculations, however,
used static (infinitely massive) scatterers. 
A simple kinematical calculation reveals that in the limit in which the energy
of the jet is much larger than the energy of the target, any angular 
distribution of radiation in the center of mass frame leads to a 
forwardly peaked distribution in the rest frame of the target.
In spite of this, even if one could argue that the scatterers were heavy,
the same calculations presented in \cite{Vitev} cannot reproduce a double peak
structure. In fact, the gluon 
distribution obtained in \cite{Vitev} are such that after integrating over the (flat) rapidity 
distribution of the associated jet, any double peak structure would disappear
in the final correlation function.
 
\emph{Deflection of the jet due to transverse expansion.} This idea has been suggested
in \cite{Fries} and is similar to those in \cite{ASW}. In this scenario, as 
the jet travels through the medium it changes its trajectory according to 
the direction of expansion of the medium it travels through. Thus, depending on the formation
point of the jet, it will be deflected in a different direction with respect to
the original axis (set by the triggered particle). Even though there are not 
quantitative estimates of this effect, we can establish one fundamental 
difference with respect to our scenario: particles are not produced in a cone around $\pi$, as
jets are deflected in either direction with respect to the original one. 
Thus, in an event by event basis, the two particle correlations should peak
either at angles smaller or bigger than $\pi$, and is the convolutions of all events which are the responsible for the double peak. This picture should be easily 
distinguished from ours via three particle correlations \cite{wolf}.

\section{ Summary: }
 We suggest that the lost energy of the
quenched jets is not just absorbed by the heat bath, but appears
in the form of hydrodynamical collective motion similar to known
 ``sonic booms'' in the atmosphere behind the
supersonic jets. The  reason for that is that QGP seems to be
a near-perfect liquid, with remarkably small dissipative effects
and robust collective flows. 
 We argue that this should result in a cone of
particles moving in the Mach direction (\ref{eqn_Mach}). 
The first data on the particle
distribution associated with the quenched jet
(the away side from high energy hadron trigger)
indeed show two peaks, with cone angles which
 agree well with our prediction.   

\ack

We thank Stony Brook PHENIX group and especially
B.Jacak for multiple discussions of the jet-related correlations
which led us to this work.
One of us (ES) thanks K.Filimonov for enlightening discussion
at its early stage. We thank H.Stocker who pointed out to us 
his talks in which similar suggestions have been made. 
This work was partially supported by the US-DOE grants DE-FG02-88ER40388
and DE-FG03-97ER4014.


\begin{thebibliography}{99}
\bibitem{fncs} Since the velocity of the shock depends on its intensity, the cone should in fact be somewhat rounded near its top. This effect is ignored in the figure.
\bibitem{fnRM} A charge/current distribution in the region near the jet
is discussed in linear response theory
in a recent paper by Ruppert and Muller \cite{RM}. Gluons and
plasmons,
with the dispersion laws suggested by pQCD at high T and lattice data,
do $not$ generate Mach cones. We thus see no ground for the
suggestion made by these authors
\end{thebibliography}
\end{document}